\documentclass[%
mathleft,%
]{an}
\usepackage{natbib}
\usepackage{times}
\usepackage{amsmath}
\usepackage{scalerel}
\usepackage{tikz}
\usetikzlibrary{svg.path}

\definecolor{orcidlogocol}{HTML}{A6CE39}
\tikzset{
  orcidlogo/.pic={
    \fill[orcidlogocol] svg{M256,128c0,70.7-57.3,128-128,128C57.3,256,0,198.7,0,128C0,57.3,57.3,0,128,0C198.7,0,256,57.3,256,128z};
    \fill[white] svg{M86.3,186.2H70.9V79.1h15.4v48.4V186.2z}
                 svg{M108.9,79.1h41.6c39.6,0,57,28.3,57,53.6c0,27.5-21.5,53.6-56.8,53.6h-41.8V79.1z M124.3,172.4h24.5c34.9,0,42.9-26.5,42.9-39.7c0-21.5-13.7-39.7-43.7-39.7h-23.7V172.4z}
                 svg{M88.7,56.8c0,5.5-4.5,10.1-10.1,10.1c-5.6,0-10.1-4.6-10.1-10.1c0-5.6,4.5-10.1,10.1-10.1C84.2,46.7,88.7,51.3,88.7,56.8z};
  }
}
\newcommand\orcidicon[1]{\href{https://orcid.org/#1}{\mbox{\scalerel*{
\begin{tikzpicture}[yscale=-1,transform shape]
\pic{orcidlogo};
\end{tikzpicture}
}{|}}}}
\usepackage{hyperref} 

\def\ms{\ifmmode {\rm M_{\odot}} \else ${\rm M_{\odot}}$\fi}    
\def\ms{\ifmmode M_{\odot} \else $M_{\odot}$\fi}    
\def\be{\begin{equation}}
\def\ee{\end{equation}}

\begin{document}

\Pagespan{1}{}
\Yearpublication{2011}%
\Yearsubmission{2011}%
\Month{1}%
\Volume{999}%
\Issue{92}%

\title{Exploring the Formation Mechanisms of Double Neutron Star Systems: An Analytical Perspective}

\author{Ali Taani \orcidicon{0000-0002-1558-1472}\inst{1}
\and Mohammed Abu-Saleem \orcidicon{0000-0002-6399-2452}\inst{2}
\and Mohammad \ Mardini \orcidicon{0000-0001-9178-3992}\inst{3,4}
\and Hussam Aljboor \inst{5}
\and Mohammad Tayem \inst{1}
}

\titlerunning{Exploring the Formation Mechanisms of Double Neutron Star Systems: An Analytical Perspective}
\authorrunning{Taani et al.}
\institute{
Physics Department, Faculty of Science, Al-Balqa Applied University,  Al-Salt 19117, Jordan
\and
Mathematics Department, Faculty of Science, Al-Balqa Applied
University, Salt 19117, Jordan
\and
Department of Physics, Zarqa University, Zarqa 13110, Jordan
\and
Department of Physics and Kavli Institute for Astrophysics and Space Research, Massachusetts Institute of Technology (MIT),
Cambridge, MA 02139, USA
\and
Department of Basic Scientific Sciences, Prince Hussein Bin Abdullah II Academy for Civil
Protection, Al Balqa applied university, Amman, Jordan}



\abstract{%
Double Neutron Stars (DNSs) are unique probes to study various aspects of modern astrophysics. Recent discoveries have confirmed direct connections between DNSs and supernova explosions. This
provides valuable information about the evolutionary history of these systems, especially regarding whether the second-born Neutron Star (NS) originated from either a Core-Collapse ($CC$) or Electron-Capture Supernovae ($ECSNe$) event. The provided scale diagram  illustrates the distribution of different types of DNSs on the basis of their orbital parameters and other factors, including mass loss. As a result, the physical
processes  in DNSs vary depending on the formation mechanisms of the second-born NS and characteristics of the systems.  $ECSNe$ processes are typically associated with merging systems ($e\times{P_{orb}}< 0.05$), while $CC$ processes are more commonly linked to non-merging systems ($e\times{P_{orb}}> 0.05$). Our results suggest a critical mass threshold of 1.30$\ms\pm0.22\ms$ (critical value) for the $ECSNe$ process to form an NS, while $CC$ processes might occur at higher masses. Examining the  orbital parameters of DNSs in a known gravitational potential can enhance our understanding of the theoretical predictions for DNS progenitor characteristics. It turns out that the $ECSNe$ process predominantly produces DNS systems with short orbital ($P_{orb} \leq 0.25 d$), nearly circular orbits ($e\simeq 0.2$), accompanied by  minimal kick velocities imparted on the proto-NS and significant mass loss.   In contrast, their orbital dynamics in a known gravitational potential plays a crucial role  in enhancing our understanding of the SNe geometry and the formation and evolution processes among different NS samples.}

\maketitle

\section{Introduction}
Chronologically, double neutron star (DNS) systems represent the endpoint of the evolution of two massive main-sequence stars. However, DNSs are formed by distinct methods in both electron-capture supernovae ($ECSNe$) and core-collapse ($CC$) events \citep{2006MNRAS.373L..50S,2006ApJ...639.1007W,2017ApJ...846..170T, 2023pbse.book.....T, 2017ApJ...835..185Y}. In the case of $ECSNe$, a system composed of two massive main sequence stars may form a $ONeMg$ core after burning of carbon \citep{2005MNRAS.361.1243P}.
The $ONeMg$ core has an upper mass of 1.37 \ms and a lower He core mass ($\sim$2 \ms). 
$CC$ SNe in massive stars (ranging from 12-25\ms) with iron cores cause implosions \citep{2010ApJ...721.1689W, 2023Galax..11...44T}, generating significant energy that compresses the core, often  leading to the formation of DNSs. Further detailed discussions can be found in Sec. 3, and their physical parameters in Table~\ref{table:1}. Researchers pay particular attention to DNSs; because  a notable fraction of them are detected as radio pulsars \citep{2006MNRAS.368.1742D, 2021ApJ...909L..19G, 2020ApJ...902L..12Z}. In addition, DNSs play a crucial role in the detection of gravitational waves, their collision rates and  NS interiors \citep{2015MNRAS.448..928K, 2010NewAR..54..140V,  2016PhRvL.116x1103A, 2017ApJ...846..170T, 2016CQGra..33h5003C}, also are thought to be the primary drivers behind the majority of short-Gamma bursts \citep{2024ApJ...970...90D}. PSR J0737-3039AB, a binary system with a recycled NS, offers  a valuable source for conducting extraordinary investigations into fundamental physics, indicating a possible (recycled) formation history from a close helium star in an NS binary \citep[e.g.,][]{2003Natur.426..531B,2004Sci...303.1153L}.  Furthermore, DNSs offer the opportunity  to place observational constraints on some fundamental scientific aspects, such as  empirical rate estimates, including the phase of mass
transfer through Roche-lobe overflow at a high rate. \citet{2015MNRAS.448..928K} argue that the non-recycled NS binary J1906 + 0746 significantly constrains the uncertainties associated with the empirical rate estimates by increasing the Galactic DNS merger rate by a factor of two. In general, DNSs have been demonstrated to be marvelous systems for the study of Einstein’s theory of general relativity,  the prediction of gravitational-waves and the strong-field regime through long-term timing observations \citep{2017PhRvL.119p1101A, PhysRevX.11.041050, 2021arXiv211106991M, 2021arXiv211106990K, 2024PhRvD.110h3040B}.

To justify the classification of SNe into CC and ECSNe types, we present several key criteria based on empirical evidence and theoretical frameworks. First, we specify the progenitor mass ranges: $CC$ typically arise from  stars with ZAMS masses  $12-25 \ms$ \citep{2006MNRAS.368.1742D, 2017ChPhL..34l9701Y}, while the $ECSNe$ progenitor lie in the range of approximately $8-12\ms$\cite{2010NewAR..54..140V}. Second, we  outline the distinct evolutionary pathways for each type: $CC$ result from the formation of an iron core, whereas $ECSNe$ originated from electron-capture processes occurring in the degenerate ONeMg cores of intermediate-mass stars \cite{2015MNRAS.451.2123T}. Additionally, we  incorporate observational evidence supporting the existence of these supernova classes, including statistics on the mass distributions and physical properties of the NSs formed from each SNe type. We  also highlight the key characteristics of these NSs, noting differences in mass, spin period, and magnetic field strength. Finally, we  present a statistical analysis demonstrating the significance of these distinctions, reinforcing the classification is grounded in empirical data and theoretical understanding rather than arbitrary assumptions.

The kick velocity of an NS is caused by the ejection of asymmetric material during the explosion of SNe \citep{2022PASA...39...40T,2022JHEAp..35...83T, 2004ApJ...612.1044P}. Typically, these velocities range between 100-500 km/s, although it can reach up to 1000 km/s in some cases for $CC$ \citep{2004ApJ...612.1044P}. $ECSNe$ process mainly lead to DNS with tight, circular orbits, low kick velocity due to the near-spherical symmetry of the explosion and lower mass ejection \citep{2010ApJ...721.1689W}. 
In addition, the $ECSNe$ process could be particularly important in producing compact binaries that include a NS \citep{2022MNRAS.513.1317Z, 2018MNRAS.481.1908K}. DNSs exhibit a wide range of orbital eccentricities (see Table~\ref{table:2}). The magnitude and direction of asymmetric kicks present uncertainties in stellar population synthesis. More models are required to characterize all binary parameters. The loss of angular momentum due to orbital decay leads to a process known an in spiral \citep{2017A&A...606A..45D}, increasing the merger event rates of DNS, making them one of the best-characterized sources of gravitational wave  for multi-messenger observations \citep[e.g.,][]{2017Sci...358.1574S, 2021AIPA...11a5309A, 2017PASA...34...24T, 2024PhRvD.110h3040B}. \citep{2023A&A...678A.187C} More recently, \citep{2017PhRvL.119p1101A} confirmed the connection between DNS and the short $\gamma$ ray burst by observing GW170817. The merging of DNS can be measured by gravitational waves, which can provide cosmological information, implications, distances, and even  extend to our understanding to the edge of the universe \citep[e.g.,][]{2017PhRvL.119n1101A}.

To date, over 100 gravitational wave events have been confidently reported in the  GWTC-3 and 4-OGC catalogs, following the first three observing runs (O1–O3) conducted by the LIGO–Virgo–KAGRA collaboration \cite{2023CQGra..40r5006A, 2023ApJ...946...59N}. The majority these detection involve binary black hole mergers, while a smaller fraction consists of binary neutron star and neutron star–black hole systems (\cite{2023PhRvX..13d1039A, 2025MLS&T...6a5054K}.  The LIGO and Virgo detectors have observed the first gravitational waves from two DNS systems in the events GW170817 \citep{2017PhRvL.119p1101A} and GW190425 \citep{2020ApJ...892L...3A}. However, many more DNS systems are expected to be observed through gravitational waves in the near future.
The literature on DNS classifications and their formation mechanisms is extensive and interesting. In \cite{10.1093/mnras/staa756, 2019Natur.568..469M, 2017hsn..book.1447W, Lasky_2015, 2021ASSL..461...53B, tauris2023physicsbinarystarevolution}, the researchers included the modeling of DNSs  as probes for general relativity and gravitational wave relativity. 
Interested readers are encouraged to refer to the provided references 
for a detailed review \citet[]{2006ApJ...639.1007W,2007AIPC..924..598V,2010ApJ...721.1689W,2012Ap&SS.340..147T,2012Ap&SS.341..601T, 2015ApJ...801...32A, 2017JPhCS.869a2090T, 2021ApJ...909L..19G, 2020ApJ...902L..12Z, 2021ChJPh..74...53A, Mardini2020, 2023RAA....23g5018A}
and the references therein for a detailed review.

The structure of the paper is organized as follows. We describe the observational sample, in Sect.~\ref{sample}. In Sect.~\ref{scenario}, the two evolutionary models are introduced. Sect.~\ref{parameter} we will analyze the dynamical properties of a set of DNSs. We will describe the mass of non-recycled NSs and the relationship of eccentricity and orbital period. The final goal is to discover a better classification method with respect to the companion mass and orbital parameters. Finally, we will end with some discussions and conclusions in Sect.~\ref{conclusions}.

\section{The observational sample}
\label{sample}

\begin{figure}
\centering
\includegraphics[width=0.5\textwidth]{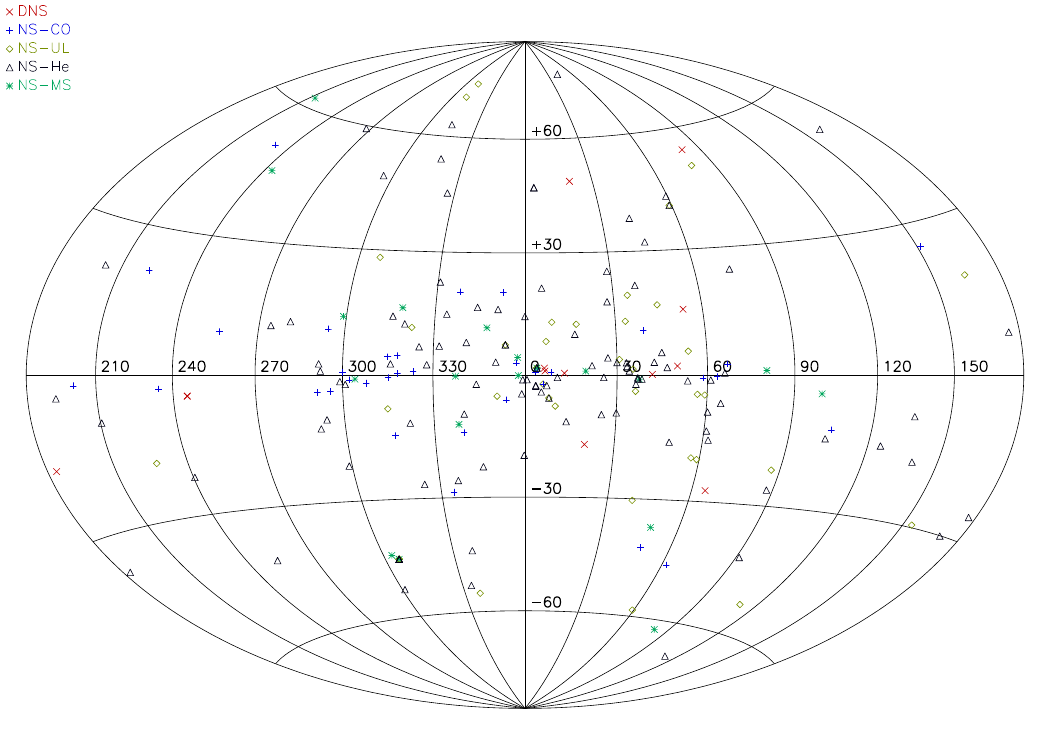}
\caption{The spatial distribution of five types of binary systems, NS-NS, NS-CO (NS and CO White Dwarf),NS-UL(NS and uncertain type star), NS-He(NS and He White Dwarf) and NS-MS(NS and  main sequence star), is depicted. The red crosses represent DNS. The pulsar data  are taken  from ATNF pulsar catalogue \citep{2005AJ....129.1993M}.}
 \label{Fig1}
 \end{figure}

\begin{figure}
\centering
\includegraphics[width=0.5\textwidth]{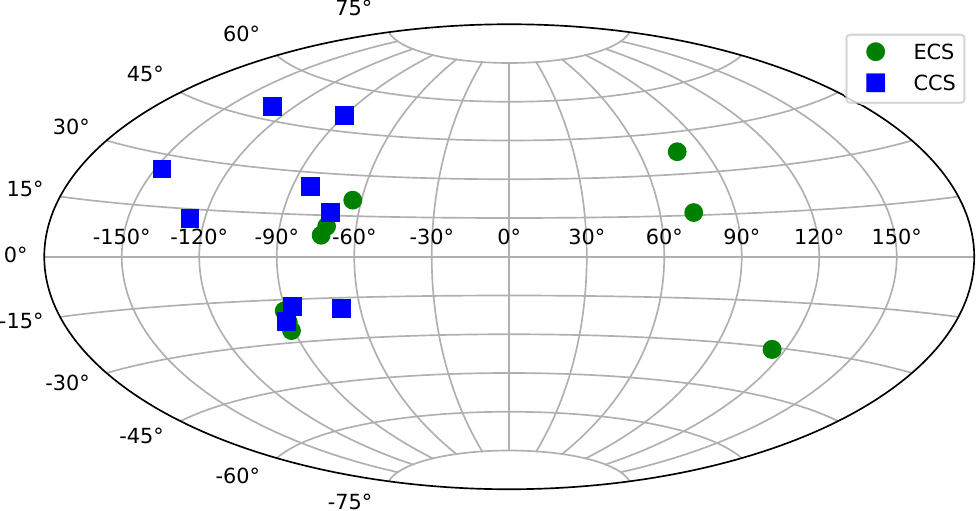}
\caption{The projected distribution components in our sample of DNS. Further observations in future missions (e.g., LOFAR \cite{2013A&A...556A...2V} and FAST) may potentially result in additional discoveries of relativistic DNSs.}
\label{projection}
\end{figure}

The DNS  systems observed  in our galaxy include 24 that reside in the galactic disk and the remaining eight in globular clusters (B2127+11C in NGC 7078, J1807-2500B in NGC 6544, J2140-2311B in NGC 7099, J1835-3259A in NGC 6652, J1748-2021B in NGC 6440, J1748-2446ao in Terzan 5, J0514-4002E in NGC 1851 and J0514-4002A in NGC 1851). We have collected relevant published data on DNS residing in the Galactic disk, as shown in Table ~\ref{table:2}. Fig.~\ref{Fig1} shows the spatial distribution of the current sample of all known DNS systems throughout the Galaxy. The scale-height of the systems is somewhat complex and uniformly distributed. The map of spatial distribution is depicted in Fig.~\ref{Fig1}, which will aid the preliminary understanding of the distribution of the Electron-Capture Supernovae ($ECSNe$) or Core-Collapse ($CC$), as we can see in Fig.~\ref{projection}.

The detailed analysis of these pulse profiles and their geometric structures was performed by \citet{2004ApJ...612.1044P, 2005MNRAS.361.1243P, 2017ChPhL..34l9701Y, 2017ApJ...846..170T}. They found that the supernova formed by the second pulsar was relatively symmetric with low mass-loss. They also suggested that,  candidates are probably forming in an $ECSNe$, or possibly the collapse of a low-mass iron core. 

In the case of PSR J1906+0746, the non-recycled companion suggests that it could be another (as in the DNS system) or a massive white dwarf (as in the NS-WD system). However, the low mass (1.29\ms) of the companion, together with its young age (113 Kyr) and magnetic field ($10^{12}$ G), makes it difficult to detect radio emission from the companion \cite{2017ApJ...835..185Y}. It is possible that there is an unseen recycled pulsar in the system, which could explain the absence of observed radio emission. More research and analysis are needed to fully understand this issue.

A comparative analysis of the physical parameters of PSR J0737-3039B (e.g. orbit eccentricity, spin down age, and the relatively high magnetic field) suggests that this pulsar can be considered as one of the components of the young DNS binary \citep{2008AIPC..983..485K, 2017ApJ...835..185Y, 2021PhRvX..11d1050K}, forming in an $ECSNe$ \citep{2015ApJ...798..118V, 2017ApJ...835..185Y}.
Another case involves the low companion mass (1.23\ms) PSR J1756-2251, which was discovered by \citet{2005ApJ...618L.119F} and is believed to have an evolution scenario for the second NS, similar to that of PSR J0737-3039B, indicating an involvement in an $ECSNe$ \citep{2014MNRAS.443.2183F}.
The relatively small eccentricity of the PSR J0453+1559 system, compared to other relatively small kick velocities at birth \citep{2015ApJ...812..143M, 2013IJMPS..23..157C, 2016RAA....16..101T}, and the large mass asymmetry are consistent with the formation via $ECSNe$.
\citet{2016ApJ...831..150L} discovered the massive PSR J1913+1102 system with a maximum total mass (2.875\ms) and an approximately circular trajectory (e = 0.09).
Furthermore, \citet{2015MNRAS.450.2922N} discovered PSR J1755-2550, a young radio pulsar with a period of (P = 315 ms) in a tight binary system with an orbital period of ($P_{orb}$ = 9.7 d). However, the nature of this system remains unclear. More recently, further studies of this system have suggested that it is a DNS system, which may be similar to low eccentricity systems \citet{2017ApJ...846..170T}.

\begin{table}[b]
\begin{minipage}{80mm}
  \caption[]{A simple statistic of key property ranges of DNS systems.}
  \label{table:1}
  \begin{tabular}{ll}
        \hline
        \hline
	\noalign{\smallskip}
	\noalign{\smallskip}
	  Parameter ranges of $ECSNe$ NSs: & \\
           $\qquad$Progenitor mass        	  	              & $8 - 12\;M_{\odot}$ \\
           $\qquad$Mass, $M_{\rm p}$     			              & $1.25-1.68\;M_{\odot}$ \\
           
           $\qquad$Mean Mass, $M_{\rm p}$     			              & $1.39\;M_{\odot}\pm 0.22\ms$\\
           $\qquad$Spin period, $P_s$	        	  	              & $144-2773\;{\rm ms}$ \\
           $\qquad$eccentricity, $e$ 		                      & $0.085-0.181$ \\
           $\qquad$Surface dipole B-field, $B$                    & $\sim$ $1\times10^{9-10}\;{\rm G}$ \\
           \hline
        \hline
	   Parameter ranges of CC  NSs: & \\
    $\qquad$Progenitor mass        	  	              & $12 - 25\;M_{\odot}$ \\
           $\qquad$Mass, $M_{\rm p}$              		          & $1.31-1.79\;M_{\odot}$ 
           \\
           $\qquad$Mean Mass, $M_{\rm p}$              		          & $1.52M_{\odot}\pm0.15\ms$
           \\
           $\qquad$Spin period, $P$  		      		          & $22.70-185.52\;{\rm ms}$ \\
            $\qquad$eccentricity, $e$ 		                      & $0.139-0.82$ \\
           $\qquad$Surface dipole B-field, $B$                   & $\sim$ $1 \times10^{11-13}\;{\rm G}$ \\
	\noalign{\smallskip}
        \hline
        \hline
  \end{tabular}
  \end{minipage}
\end{table}

\begin{table*}\large
 \centering
 \begin{minipage}{400mm}
  \caption[]{Parameters of DNS are provided from observations of pulsars in the Milky Way.}
  \label{table:2}
  \setlength{\tabcolsep}{5pt}
  \begin{tabular}{cccccccc}
   \hline
System  &$M_p(M_{\odot})$  & $M_c(M_{\odot})$   &$B (\times10^{9} G)$     &$P_{orb}(d)$ &$P_s(ms)$  &$e$     &$References$ \\

\hline
$ECSNe$ \\
J0737-3039A/B     &$1.25$    &$1.33$      &6.1         &0.102          &22.70     &0.09    &$1,2$      \\
J1906+0746     &$1.29$    &$1.32$      &1700            &0.17          &144.07     &0.09        &3     \\
J1756-2251     &$1.34$     &$1.23$         &5.5        &0.32         &28.46      &0.18        &4      \\
J0453+1559     &$1.56$     &$1.17$       &2.9           &4.07           &45.78      &0.11        &5     \\
J1913+1102     &    1.62       &   1.27            &2.1         &0.21          &27.29      &0.09        &6       \\
J1755-2550     &     1.3    &     $>0.4$      &  270               &9.7          &315.2       &0.09        &7        \\
J1946+2052     &$1.25$    &$1.25$           &4    &0.09          &17     &0.06    &8     \\
J1757-1854     &$1.34$    &$1.40$          &7.6     &0.18          &21.5     &0.61    &9      \\
J0509 +3801     &$1.34$    &$1.46$         &4.3     &0.38          &76.54     &0.59    &10      \\
J1325-6253   & $1.59$& $0.98$   & 1.19 & 1.815 & 22.96&   0.064 & 11\\
 J1018-1523 & $1.2$ & $1.1$ & 3 & 8.9 & 83 & 0.23 & 12 \\
 J1155-6529& 1.4 &1.27&----&3.76&79&0.26&13\\
 J1208-5936&1.26&1.32&30&0.63&78.7&0.35&14\\
 B1913+16      &$1.44$    &$1.39$       &2.3 &0.32          &59.03      &0.62        &15      \\
 B1534+12      &$1.33$    &$1.35$       &9.7         &0.42          &37.90     &0.27         &16     \\
  J1901+065&1.68&1.1&4.1&14.45&75.7 &0.366&17\\
\hline
CC SNe \\

J1518+4904    &$1.42^{+0.51}_{-0.58}$     &$1.29^{+0.58}_{-0.51}$     &1.07 &8.63    &40.94    &0.25     &18     \\
J1829+2456    &1.31             &1.30                 &1.5    &1.18          &41.01      &0.14        &19    \\
J1930-1852    &$\le$1.32      &$\ge$1.30         &58          &45.06          &185.52      &0.40        &20     \\
J1811-1736    &$<1.64$            &$>0.93$        &9.8      &18.80         &104.2       &0.82         &21      \\
J1753-2240    &           &$>0.49$             & 9.7         &13.64         &95.14      &0.30        &22    \\
J1411+2551 & $<1.62$& $>0.92$      &5.45 &0.4       & 0.06& 0.17&23\\
J1759+5036&1.79&0.84&9.5&2.04&176&0.30&24\\
J2150+3427 & $<1.67$ & $>0.98$ & 49& 10.6 & 654&0.6&25\\

\hline
\hline

\end{tabular}
\scriptsize\\
($1$)\citet{2004Sci...303.1153L,2006Sci...314...97K,2013ApJ...767...85F};
($3$)\citet{2015ApJ...798..118V};
($4$)\citet{2005ApJ...618L.119F,2008A&A...490..753J};\\
($5$)\citet{2015ApJ...812..143M};
($6$)\citet{2016ApJ...831..150L};
($7$)\citet{2020Natur.583..211F};
($8$)\citet{2015MNRAS.450.2922N};
($9$)\citet{2018ApJ...854L..22S};
($10$)\citet{2018MNRAS.475L..57C};
($11$)\citet{2018ApJ...859...93L};\\
($12$)\citet{2022MNRAS.512.5782S};
($13$)\citet{2023ApJ...944..154S};
($14$)\citet{2023MNRAS.524.1291P};
($15$)\citet{2023A&A...678A.187C};
($14$)\citet{2014ApJ...787...82F};
($15$)\citet{1975ApJ...195L..51H,2010ApJ...722.1030W,2016ApJ...829...55W};
($16$)\citet{2014MNRAS.443.2183F};
($17$)\citet{2004MNRAS.350L..61C};\\
($18$)\citet{2015ApJ...805..156S};
($19$)\citet{2007A&A...462..703C};
($20$)\citet{2009MNRAS.393..623K};
($21$)\citet{2017ApJ...851L..29M};
($22$)\citet{2021ApJ...922...35A};
($23$)\citet{2023ApJ...958L..17W};
($24$)\citet{2024MNRAS.530.1506S};
 \end{minipage}
\end{table*}

Table~\ref{table:2} lists detailed fundamental parameters  for known DNSs-- including pulsar ID, mass of recycled pulsar ($M_p$), mass of non-recycled pulsar ($M_c$),  magnetic field ($B$), orbital period($P_{orb}$), spin period ($P_s$), eccentricity ($e$), and the corresponding references. As we mentioned in the table, we divided the DNSs into two groups based on eccentricity and orbital period. A specific criterion $e~x~{P_{orb}} = 0.05$ is used to distinguish between these classes. In contrast to systems with $e~x~{P_{orb}} < 0.05$ falling in the region of indicative of merging systems (see \citep{2020Natur.583..211F}, they exhibit low eccentricity and short orbital periods, suggesting that they were formed by the ECSNe process. Systems a $e~x~{P_{orb}} > 0.05$ belong to a distinct region with a large eccentricity and long orbital periods, reflecting more eccentric orbits due to the significant velocities imparted during the explosion (see \citep{2004ApJ...612.1044P}.  It shows that it forms by CC processes. This classification is determined by the supernova kick velocity and the nature of the binary system partners \citep{2006MNRAS.373L..50S, 2017ApJ...835..185Y, 2017ApJ...846..170T, 2023pbse.book.....T}. It should be noted that among the eight   pairs of DNS (B1534+12, B1913+16, J1518+4904, J1829+2456, J1930$-$1852, J1811$-$1736, 
J1753$-$2240 and J1411$-$2551), the NS in these systems is believed to  have formed through core collapse \citep{2016ApJ...829...55W,2018ApJ...867..124S}. This conclusion is  based on their relatively long orbital periods with high eccentricities  and a companions mass of approximately  1.30$\ms\pm0.22\ms$.

\section{DNS Evolutionary Scenario}
\label{scenario}

Numerous authors have thoroughly studied the formation of DNS systems \citep{2006MNRAS.373L..50S,2006ApJ...639.1007W, 2010ApJ...721.1689W,2017ApJ...846..170T, 2024ApJ...970...90D, 2024MNRAS.530.1506S}. However, the core collapse process is commonly referred to in the literature as the standard scenario. This is because it is believed to be the predominant  formation mechanism for DNSs, where both components of the DNS have masses ranging from approximately  $\sim 12\ms-25 \ms$ \citep{1991PhR...203....1B}.

\begin{figure}[ht]
   \centering
  \includegraphics[width=0.5\textwidth]{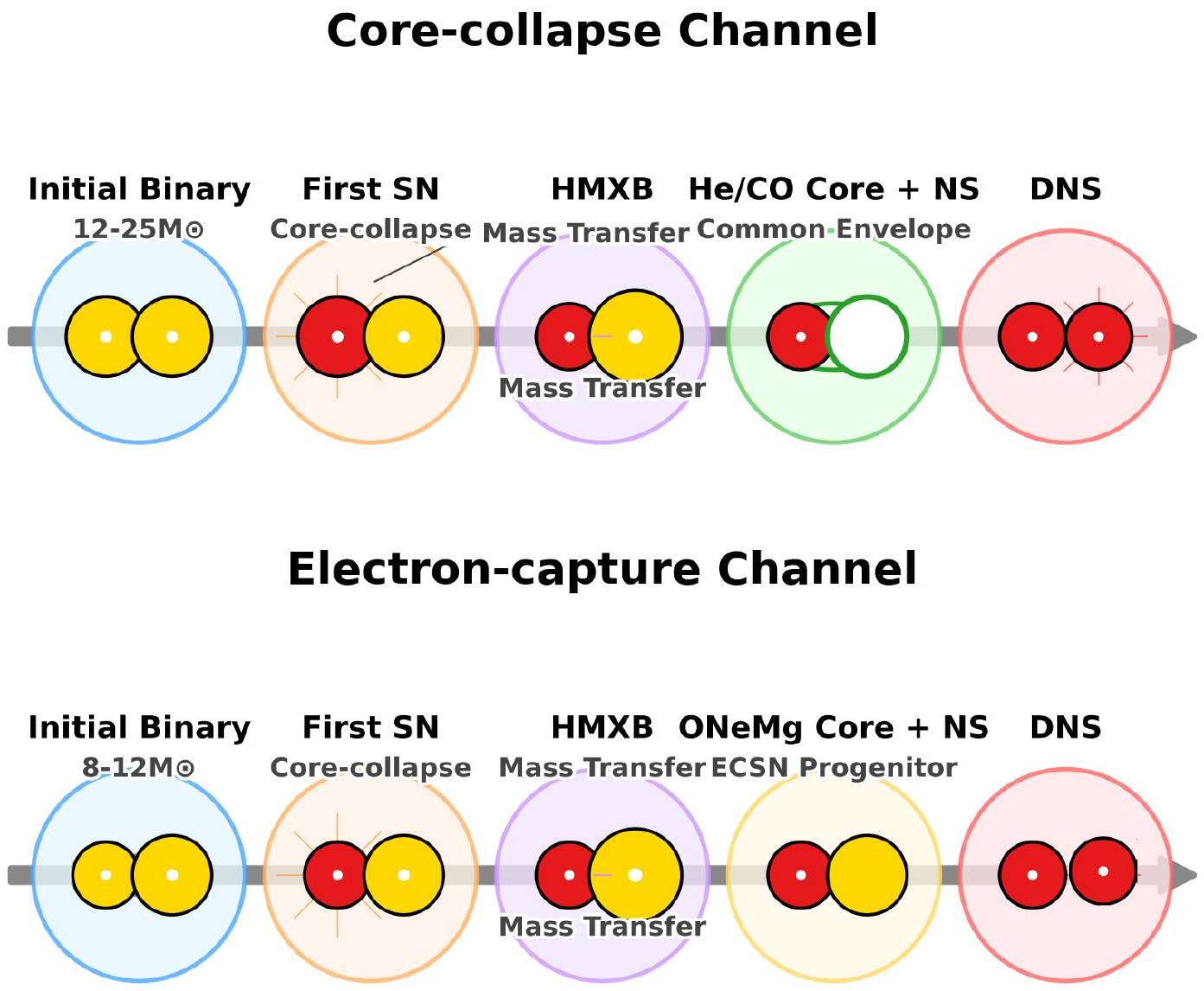}
   \caption{The formation sketch map of DNS binaries in the Galactic disk is shown in the two panels. The \emph{Top-panel}.
    illustrates the  $CC$ scenario, where the system demonstrates considerable proper motion, indicating that its formation might be associated with a $CC$ event. This motion could have arisen from the explosion, imparting significant velocity to the system. The \emph{Bottom-panel} illustrates the  $ECSNe$ scenario, where the observed trend of the companion NS mass continually falling below that of the recycled NS suggests the possibility of an $ECSNe$ scenario.}
   \label{Fig2}
   \end{figure}

The Figure \ref{Fig2} presents the two evolutionary  pathways for the formation of DNSs. The $Top-panel$ of Figure \ref{Fig2} illustrates the formation process of the $CC$ scenario. This scenario can be summarized as follows:\\
(i) At the start, the system consists of two rotating main-sequence stars, designated  as Star 1 and Star 2, initially having masses ranging from 12\ms-25\ms.\\
(ii) The first NS forms in a HMXB\citep{1999PhRvL..83.3776L, 2015MNRAS.451.2123T} following first supernova event.\\
(iii) The current system consists of a NS and  CO or He core. Due to angular momentum and dynamical friction, there is an exchange of mass, either  stably or unstable, within a common envelope (CE) between these components \citep{2019AstBu..74..464M}.
This process eventually lead to the  second SNe explosion of the CO or He core\citep{2007AIPC..924..598V}. A system demonstrating considerable proper motion \citep{2013ApJ...767...85F}, might indicate its formation through $CC$.
This motion could have been caused by the explosion, which imparted significant velocity to the system.
Moreover,  if the mass of the recycled NS is higher than that of its companion, it implies formation through the $CC$.

The $Bottom-panel$ of Figure \ref{Fig2} illustrates the formation of DNS through the $ECSNe$.
The $ECSNe$ scenario was originally described by \cite{1980PASJ...32..303M} and \cite{1987ApJ...322..206N}.
In this scenario:
The formation process of $ECSNe$ scenario is summarized as follows:\\
(i) A system is made up of two rotating 8\ms -12\ms main sequence stars.
(ii) After the first SNe, the formation of the first NS in a LMXB \citep{2004ApJ...612.1044P, 2005A&A...435..231G};\\
(iii) The current system consists of a NS and a strongly degenerate $ONeMg$ core. Burning the external shell of $ONeMg$ core will result in a mass increase to the Chandrasekhar limit. When $^{24}Mg$ and $^{24}Ne$ will capture electrons, the electron degeneracy pressure will be reduced \citep{2019ApJ...875...89M}, resulting in a core that  reaches the nuclear matter upper limit density ($\sim 4.5\times10^9 g/cm^3$) in the rapid contraction process. A lower mass NS will be formed after in  next  SNe.
The mass of the companion NS remains consistently lower than that of the recycled NS, strongly indicating the likelihood that $ECSNe$ \citep{2017ApJ...835..185Y}.

\begin{figure}[ht]
\centering
\includegraphics[width=8.0cm, angle=0] {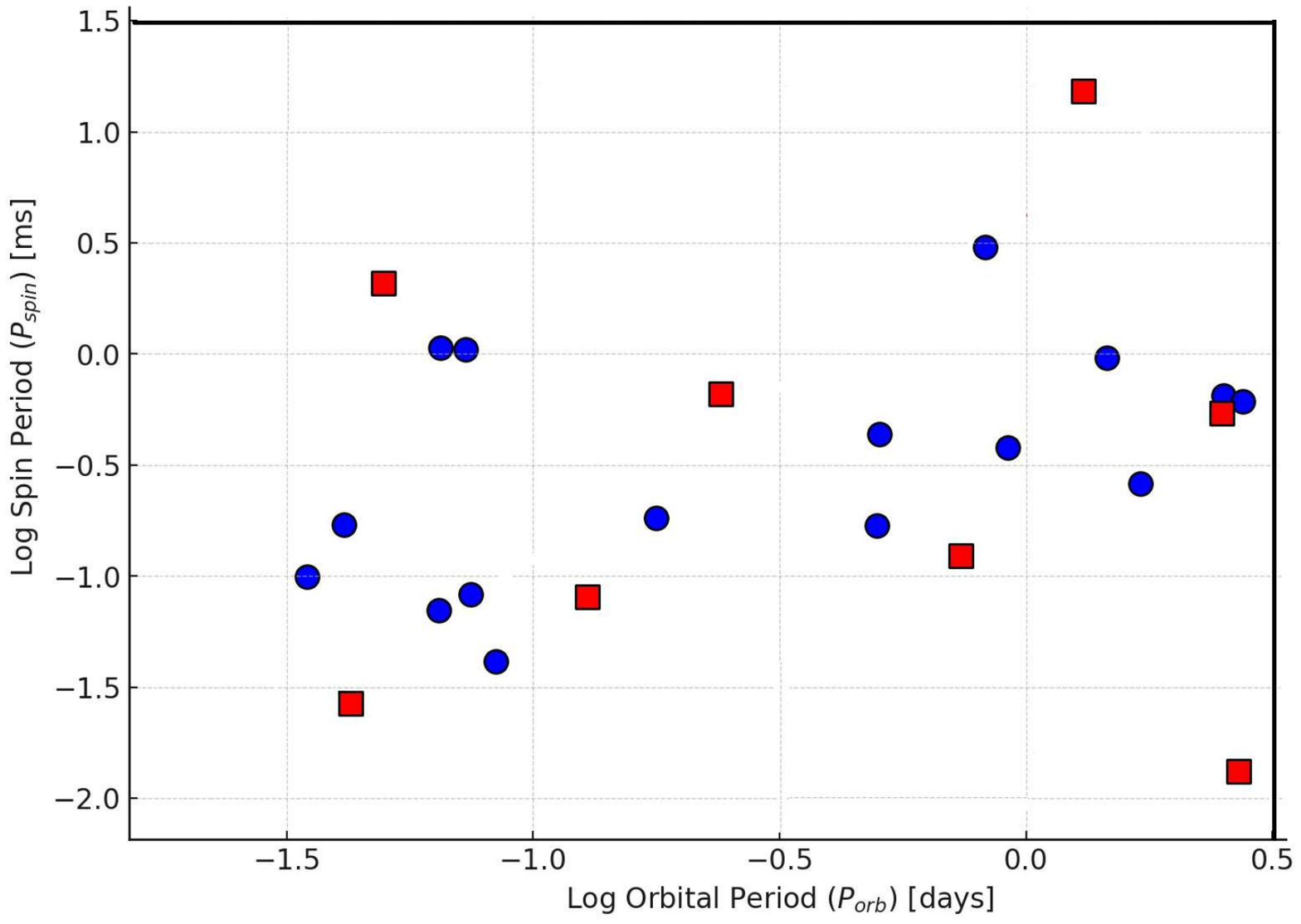}
\caption{The schematic depicts the approximate location of DNSs sample via the $ECSNe$ (white circles) and $CC$ (black squares) in the P$_{spin}$ - P$_{orb}$ diagram. }
\label{Fig-new.pdf}
\end{figure}

The criteria presented in Table~\ref{table:1} serve to classify DNSs into distinct groups based on unique properties, show the differences in their attributes and connected to the two different formation mechanisms. This classification is based on the unique characteristics of each mechanism. Finally,  a note should be made for the Corbet diagram  \citep{1986MNRAS.220.1047C} concerning the specific combination between merging and non-merging systems with different formation pathways and characteristics of the systems,  $CC$ and $ECSNe$.   As we can see in Fig.\ref{Fig-new.pdf},  that all the second-born NS binaries in DNSs forming in a $ECSNe$, are thought to experience a merging phase and then produce gravitational-wave observations, except for the system (B1534+12) \citep{2014ApJ...787...82F}. On the other hand, systems formed via the $CC$ are thought to experience a non-merging phase, except for the two systems (J0453+1559 and J1755-255) \citep{2015ApJ...812..143M}. Future detection by LIGO and VIRGO will  reveal much information about these systems. This result agrees with recent work by \citet[]{2020ApJ...902L..12Z}.

\section{Parameter analysis of DNS}
\label{parameter}

This section discusses the eccentricity, orbital period, and mass of the second-born NS, to better distinguish between the DNS category.

\begin{figure}[ht]
\centering
\includegraphics[width=8.0cm, angle=0] {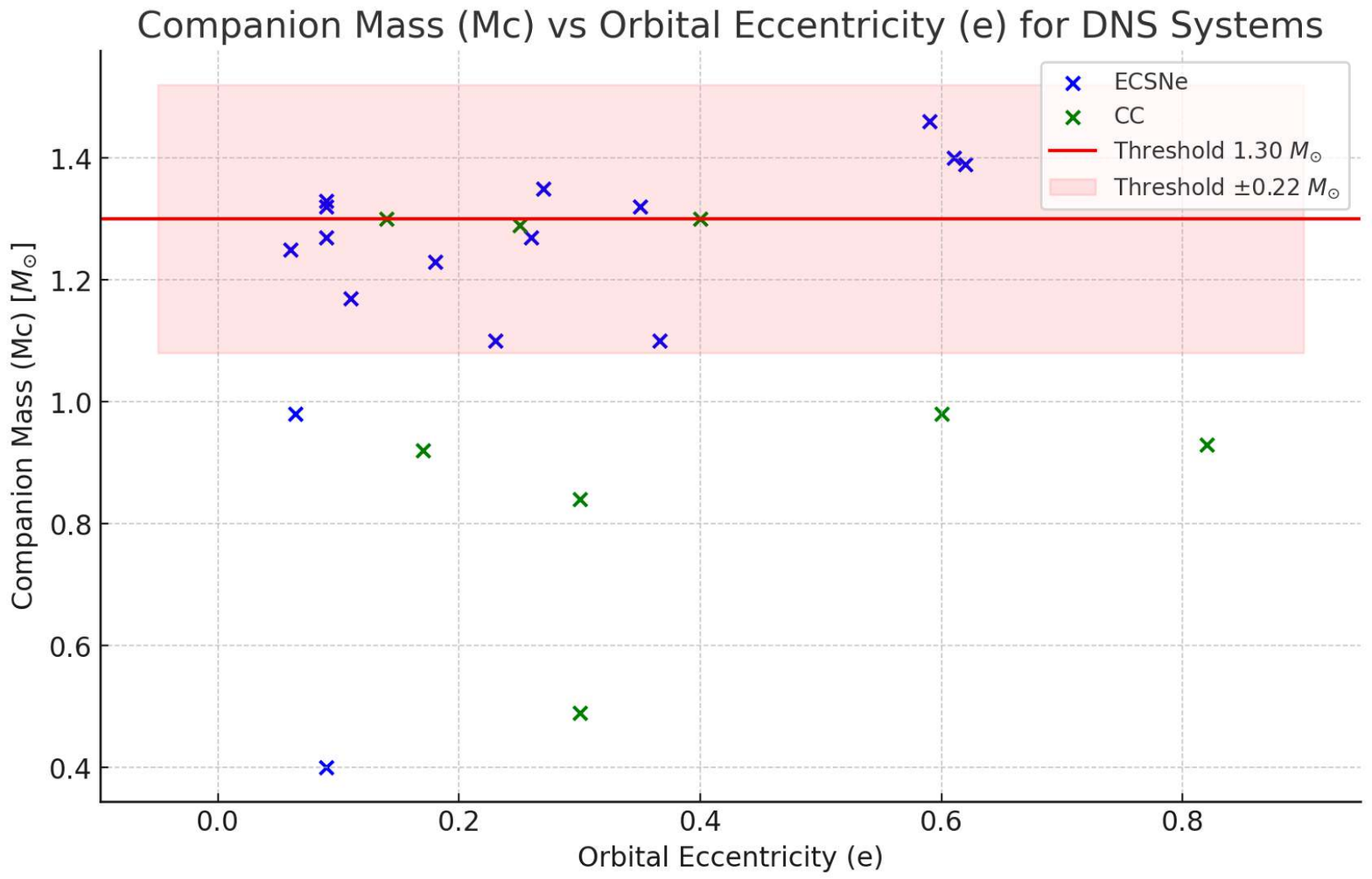}
\caption{The diagram plots the companion mass (M$_c$) against the orbital eccentricity(e) for DNSs. The red line marks the threshold at 1.30$\ms$, with the shaded area the indicating the $\pm0.22\ms$ range. This threshold visually differentiates systems likely formed via  $ECSNe$ and those formed through  $CC$ events.} 
\label{Fig3}
\end{figure}

As shown in Table \ref{table:2}, most companion masses are accurately measured and collected from the literature. However, the lower limit of the companion mass is given for four systems (PSR J1829+2456, J1930$-$1852, J1811$-$1736 and J1753$-$2240). Here, we do not consider the DNS systems residing in globular clusters (eight in total); since their characteristic properties are uniquely different  from those within the Galactic disk (see Fig.\ref{Fig3}).

\begin{figure}
\includegraphics[width=9.0cm, angle=0] {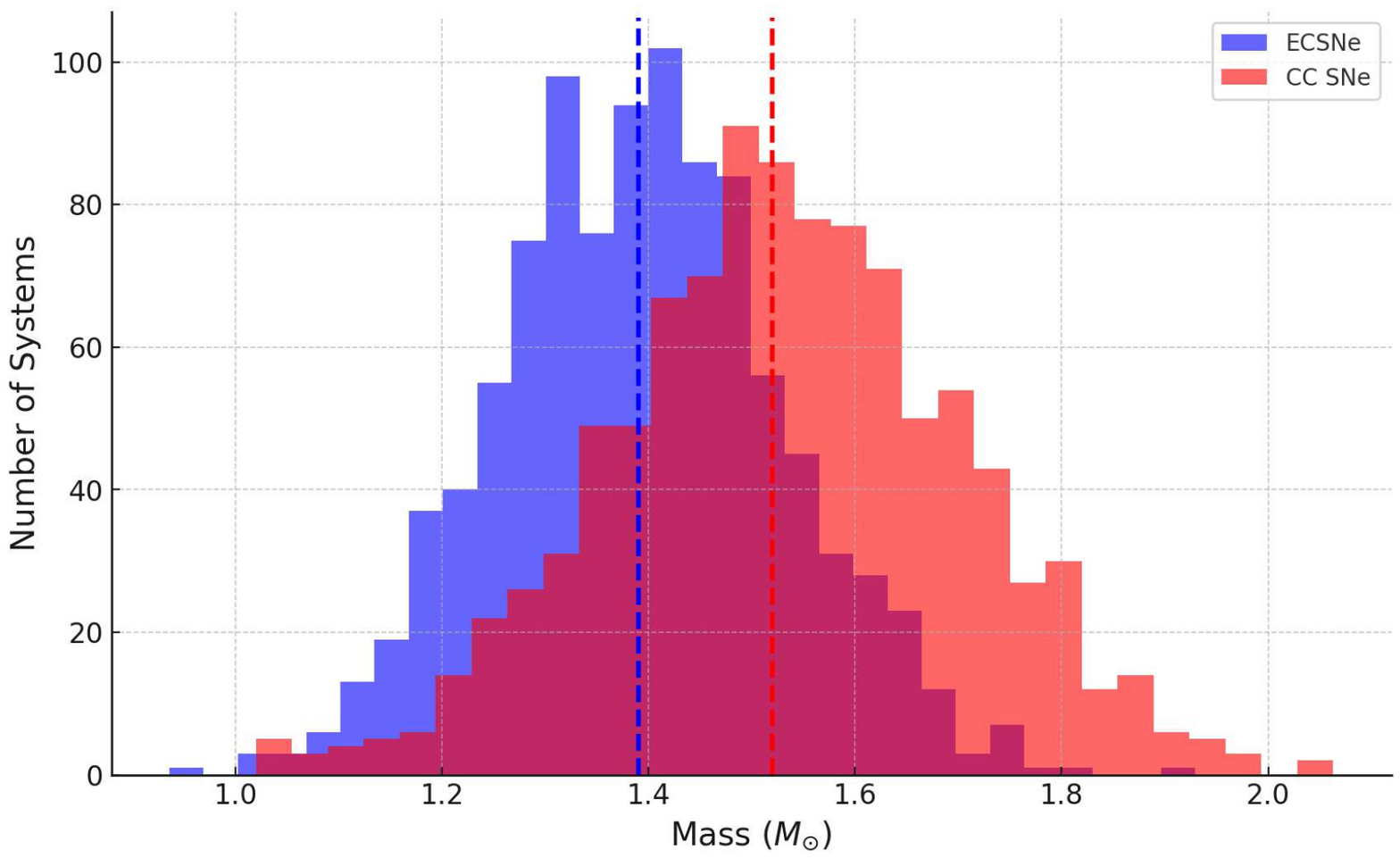}
\caption{The histogram illustrates the mass distribution of the DNS systems formed via  $ECSNe$ (with a mean mass of 1.39\ms) and $CC$ (with a mean mass of 1.52\ms). The dashed vertical lines indicate the average mass of each group. }This visual representation highlights the difference in mass between the two formation scenarios, providing the insight into characteristics  and evolutionary processes of these binary systems.
\label{Mass}
\end{figure}

The masses of the 24 DNSs listed in Table \ref{table:2} have been measured with high timing  accuracy \citep{2011A&A...527A..83Z, 2016ARA&A..54..401O}. It is noticeable that the masses of all recycled NSs born and spun-up by accreting matter from their companion star~\citep{1991PhR...203....1B,1994ARA&A..32..591P,1995JApA...16..255V,2006MNRAS.366..137Z}, falls within a stellar mass range of 1.29\ms-- 1.56\ms. In contrast, the mass of the second-born NS and non-recycled ranges from 1.17\ms to 1.39\ms. Consequently, it is evident  that the average stellar mass  of non-recycled NS is higher than that of the recycled ones. The mass distribution of DNSs has a narrow deviation of approximately 0.02\ms. Half of the  DNSs listed in Table \ref{table:1} exhibit a narrow stellar mass deviation, approximately  (1.32$\pm$0.024 M$_{\odot}$).
Fig. \ref{Mass} shows the distribution of the NS masses in DNS systems ($ECSNe$ and $CC$). The histogram divided into two groups ( recycled and non-recycled). 
Furthermore, the mass distribution of recycled NSs tends to cluster around higher values 1.48$\pm$0.025M$_{\odot}$ reflecting the mass accretion that occurs binary evolution. In addition, this figure highlights that $ECSNe$ typically produce lower mass NSs (1.37$\pm$0.023M$_{\odot}$), whereas $CC$ are associated with higher mass systems. This distinction is further supported by the analysis $T-test$, which reveals a statistically significant difference between the mass distribution of recycled and non-recycled NS, with a confidence level of \%95. The narrow mass range for DNSs suggests that the evolutionary pathways for these systems are constrained by specific physical processes and parameters during SNe, providing valuable information on the formation of NSs.

Table~\ref{table:1} lists the criteria for justifying whether a DNS is formed via $ECSNe$ or $CC$ SNe. A simple statistical overview of key property ranges for DNS systems is presented. 
The categorization of DNSs into two groups was performed on the basis of the specific criteria that reflect the unique characteristics associated with the two distinct formation mechanisms. 

The dichotomy in the magnetic field strengths between NSs formed via $ECSNe$ and $CC$  originates from their fundamentally distinct explosion mechanisms and post-formation evolution paths \citep{2012Ap&SS.340..147T}. $ECSNe$, which arise from low-mass progenitors, produce relatively symmetric, low-energy explosions that preserve the magnetic field configuration of the progenitor, typically resulting in weaker surfaces fields ($\sim$ 10$^{9-10}$ G; e.g., \citealt{2011A&A...527A..83Z}). In contrast, $CC$  involve violent, asymmetric explosions, where  extreme turbulence and convective dynamos during collapse and amplify surface fields to $\sim$ 10$^{11-13}$  G (e.g., \cite{2019A&A...622A..74J, 2012ARNPS..62..407J}). 
Furthermore, recycled NSs, which often form in close binaries, can accrete substantial mass of $\sim  $ 0.1\ms-0.2ms\ \cite{2011A&A...527A..83Z}, leading to longer lifespan and magnetic field decay due to  accretion-driven burial, as well as faster spin periods (\citealt{2012Ap&SS.340..147T}). In contrast, non-recycled NSs formed directly from $CC$ evens typically retain strong magnetic fields ($\sim$ 10$^{12}$ G) and experience shorter spin-down timescales (\citealt{2017JApA...38...45V}). This combination of explosion dynamics and accretion history fundamentally shapes the observed magnetic field distributions among NSs.




\subsection{DNS mass distribution  and T-test}
\label{Sect.2.1}

We used statistical $T-test$ to quantitatively evaluate our sample and compare the mean values of the two groups of DNS components.  This method involves defining the parameter $t$ to determine if the means (averages) of two sample groups with different types of companions are significantly different ~\citep{John2006}:

\be
\label{E1}
t = \frac{\overline{X_1}-\overline{X_2}}{S_p\sqrt{{2/n_1 + 2/n_2}}}\;,
\ee
 where  $\overline{X_1}$ and $\overline{X_2}$ are the averages of two groups,  (n1 =16), (n2 =8) are the number of samples in each group that defines the degree of freedom for this test ($d=2n-2=16$). 
 The pooled standard deviation is denoted  by the parameter $S_p$,

\be
\label{E2}
S_p = \sqrt{\frac{S^2_{X_1}+S^2_{X_2}}{2}}\;,
\ee
where $S^2_{X_1}$ and $S^2_{X_2}$ are the unbiased estimators of the variances of two groups.

Based on the updated data from the two  DNS mass groups, we  calculated  the t-parameter, yielding
 t=2.02. Our analysis further reveals that for the significance levels of 95\% and 99\%, the critical t-values are  $t_{0.05(d=24)}=2.075<t<t_{0.05(d=24)}=2.075$, and $t_{0.01(d=24)}=2.82<t<t_{0.01(d=24)}=2.82$. Since our calculated t-value falls below the 95\% threshold, the results indicate that the difference in mass distributions between the two groups (recycled NSs and their companions) suggests a potential trend where the stellar mass of the  recycled NSs may tend to be higher than that of the non-recycled companions (see Table \ref{table:2}).

When  comparing the mass distribution of DNSs with that of MSPs, we find that the deviation of the DNS mass distribution is approximately $\sim$ 0.02 \ms, which is roughly  ten times lower than that of the MSPs (approximately $\sim$0.2 \ms) as reported by \citep{2011A&A...527A..83Z}. The recent measurement  of NS mass statistics, including around  $\sim$70, indicates  that all the measured NS masses are distributed within  a broader range, approximately \citep[$\sim$ 1 \ms - 2\ms][]{2011A&A...527A..83Z, 2016ARA&A..54..401O}. Consequently, we can conclude that the specific  evolutionary history of DNSs  significantly contributes to their mass formation, and there must be  a mechanism constraining the birth mass of DNSs to a narrower range. The observed mass range reflects both the initial mass distribution and the impact of the accretion phase \cite{2004ApJ...612.1044P, 2020Natur.583..211F}.

\begin{figure}
\includegraphics[width=8.0cm, angle=0] {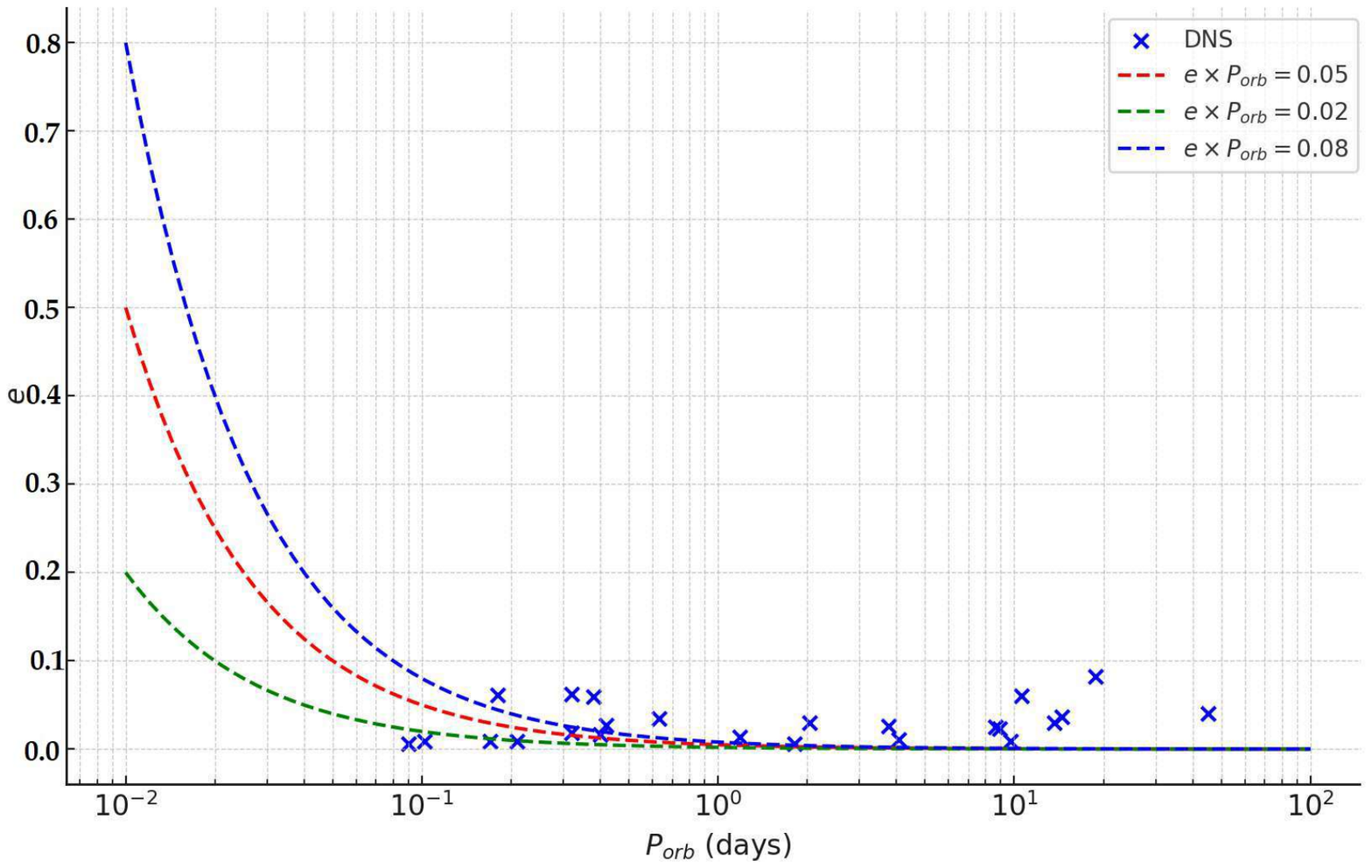}
\caption{The diagram plots the eccentricity (e) against orbital period for the DNSs.
The colored curves represent the condition  $e\times{P_{orb}} = 0.05$ (with $P_{orb}$ in unit of days). This condition separates  the DNS
samples into two regions based on their  formation pathways and system characteristics: $ECSNe$ (where $e\times{P_{orb}}< 0.05$ ) and $CC$  (where $e\times{P_{orb}}> 0.05$),
respectively.}
\label{Fig4}
\end{figure}

Fig.~\ref{Fig4} presents the fitting curves based on  condition ($e~x~{P_{orb}} = 0.05$), which serves as a boundary between DNS systems formed via $ECSNe$ and $CC$.
The region below the  line ($e~x~{P_{orb}}< 0.05$) corresponds to systems with lower eccentricities and shorter orbital periods, indicative of formation through $ECSNe$, 
where SNe imparted minimal kicks to the systems. On the other hand, the region above the line  ($e~x~{P_{orb}} > 0.05$) represents systems with higher eccentricities 
and longer orbital periods, characteristics of formation though $CC$ mechanisms, where larger kicks are expected. Several systems deviate from the curves' predictions,
indicating the presence of additional factors such as significant mass transfer or unique evolutionary pathways. More observations are needed for further investigation \citep{PhysRevX.11.041050, 2024ApJ...970...90D} 


\begin{figure}[ht]
\centering
\includegraphics[width=0.5\textwidth]{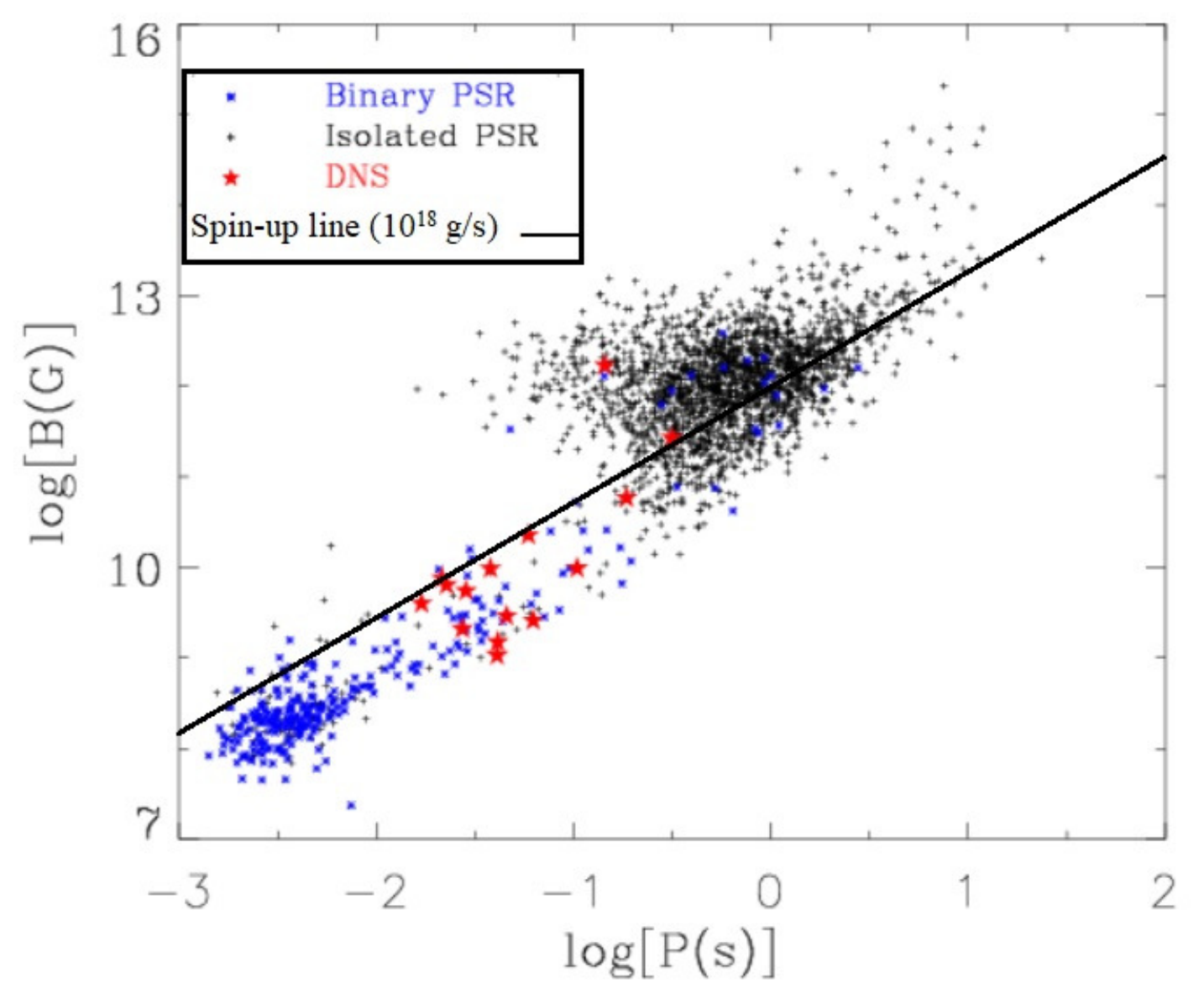}
\caption{The magnetic-field versus spin-period diagram for  DNSs is shown. The solid lines represnt the spin-up lines with the  accretion rate of $10^{18}$ g/s.  The isolated pulsars, binary pulsars and DNSs are labeled with  dots, pentagram and stars, respectively (for details see \citep{1991PhR...203....1B}). The pulsar data   in   this figure are taken  from ATNF pulsar catalogue \citep{2005AJ....129.1993M}.}
 \label{spin-up-line}
\end{figure}

To further investigate the evolution history and mass formation conditions of DNS, we compare DNSs with MSPs using the  magnetic field versus spin period (B-P) diagram, as shown  in Figure \ref{spin-up-line}. Two DNS systems, J1906+0746 (B$\sim 17 \times10^{12}$ G, P$_{s}$=144 ms) and J0737-3039B (B$\sim 16 \times10^{12}$ G, P$_{s}$=773ms) appear above the spin-up line, which is unusual for DNSs. These systems exhibit  spin periods that are too long for their magnetic fields according to spin-up theory, suggesting that they behave differently  from typical recycled pulsars, which generally fall along the spin-up line.  \citep{2017ApJ...835..185Y}, conducted an in-depth analysis of these two DNSs and concluded that both systems share a similar origin and evolutionary process, likely involving e-capture for the formation of the second-born NS. Furthermore, their companions are expected to be a long-lived recycled-pulsars.

However, we excluded PSR J1807-2500B and J1411 + 2551 because of the high uncertainty in their magnetic field. 

The evolutionary history of DNSs and MSPs is evident in their distinct positions on the B-P diagram. MSPs generally accumulate around $\geq$ 0.1\ms-0.2 \ms\, while DNSs absorb approximately  $\sim $0.001\ms-0.01 \ms\ from their companions \citep{1991PhR...203....1B, 2006MNRAS.366..137Z,2015AN....336..370P}. Furthermore, MSPs are the result of LMXBs (e.g., companion mass of $ \sim 1\ms-2 \ms$), whereas DNSs are primarily involved in HMXBs (e.g., companion mass of 8\ms-10 \ms for  electron capture \citep{1987ApJ...322..206N} and 10-25 \ms for the SN explosion \citep{2006MNRAS.368.1742D}). Generally, the LMXB is very long lived at a time scale of a dozen billion years, so the MSP has a sufficient time to  accumulate  matter from the companion. In addition, the less massive  companion star  holds a small size, it is weakly influenced by the primary star  when compared  to the case of HMXB. As a result, the mass distribution  of the MSPs inherits that of the isolated NSs by SNe explosions, by adding the accretion mass of $\sim  $ 0.1\ms-0.2 \ms\. Therefore, the distribution of  MSP masses is very broad,  from $\sim 1.2 \ms $ to $\sim  > 2.0 \ms$ \citep{2011A&A...527A..83Z, 2016ARA&A..54..401O}.

We  conducted an investigation of statistically significant similarities among these DNSs by employing unsupervised clustering algorithms, specifically \texttt{HDBSCAN}. This analysis would help us to understand whether these stars experienced similar evolution histories. Our clustering analysis focused on crucial parameters:  Mass, orbital period and magnetic field parameters. To facilitate this analysis, normalization of these variables was necessary. We achieved this normalization using the \texttt{Astropy.stats.biweight\_location()} \\ and \texttt{astropy.stats.biweight\_scale} routines \footnote{For a detailed description of the biweight\_location() algorithm visit: https://docs.astropy.org/en/stable/api/astropy.stats.biweight\_location.html}. As a result, we identify the DNSs distribution's central position as the following formula:

$\zeta_{biloc}= M + \frac{\sum_{|u_i|<1} \ (x_i - M) (1 - u_i^2)^2} {\sum_{|u_i|<1} \ (1 - u_i^2)^2}$

We also compute the standard deviation of DNSs distribution using the following formula:

$\zeta_{biscl} = \sqrt{n} \ \frac{\sqrt{\sum_{|u_i| < 1} \ (x_i - M)^2 (1 - u_i^2)^4}} {|(\sum_{|u_i| < 1} \ (1 - u_i^2) (1 - 5u_i^2))|}$

\begin{figure*}[htbp!]
\includegraphics[width=1.1\textwidth]{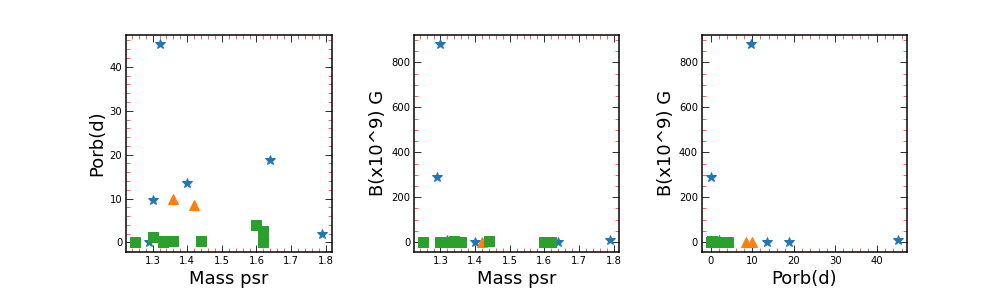}
\caption{Left panel illustrates the two identified clusters in the $M$, P$_{orb}$ space. The middle panel displays the two identified clusters in the $M$ and $B$ space. The right panel exhibits  the two identified clusters in the P$_{orb}$ and $B$ space. Asterisk symbols represent stars that do not fall into any cluster (Noise).}
\label{fig:clusters}
\end{figure*}

After normalizing the selected parameters, we utilized specific \texttt{HDBSCAN} parameters: \texttt{min\_cluster\_size} = 5, \texttt{min\_samples} = 5, \texttt{cluster\_selection\_method} = leaf, \texttt{prediction\_data} = True, and minimum confidence set to 20$\%$. These combinations of parameters led to the identification of two clusters. In Figure~\ref{fig:clusters}, we can observe  the tagged stars in each of the chosen parameters spaces. Filled pointing-up triangle and square symbols denote the first and second cluster, respectively. Asterisk symbol represent Outliers DNSs; differs significantly from other observations. By examining the true cluster labels (i.e., $ECSNe$ and $CC$) of the two groups, we found that the $ECSNe$ and $CC$ source are likely to be well separated in 3D-space (mass, orbital period, and magnetic field). Finally, considering the strong correlations and degeneracy between parameters that affect evolutionary paths of DNS formation, the application of \texttt{HDBSCAN} would require various choices to investigate to understand how the assumptions would affect the results.


-------------------

\section{Discussions and Conclusions}
\label{conclusions}

It is interesting to study DNS masses and compare them with MSP masses, as the nuclear matter compositions are different. The mass estimation of NSs with different mass-transfer histories is shown in a B-P diagram. As a result, the mass distribution of the MSPs will have a distribution similar to that of the isolated NSs. Furthermore, if further mass accretion exceeds the range of $0.1\ms -0.2 \ms$, it is possible to explain the heavy MSPs distributed in a broader regime ($\sim 1.2 \ms -   2.0 \ms$) than DNSs. This includes nuclear matter classifications, \textit{r}-process nucleosynthesis of elements \citep{2019ApJ...882...27M, 2019ApJ...875...89M,Mardini2020, 2021AN....342..625A}, gravitational waves, and neutrino emission between DNS and MSPs \citep{2019JPhCS1258a2024M}.

We applied the $T-test$ to statistically evaluate the differences in mean mass distribution between recycled and non-recycled components within the DNSs. The calculated means and standard deviations for the two samples are 1.370$\pm$0.023M$_{\odot}$ and 1.48$\pm$0.025M$_{\odot}$ respectively. A distinction of approximately ($\Delta M = ~0.1\pm0.034M_{\odot}$) with a deviation of approximately 3 sigma was observed. The application of \texttt{HDBSCAN} in our analysis enhances the robustness of our clustering approach, efficiently identifying the formation and evolution of DNSs. This method reveals significant patterns and relationships in the diverse population of DNSs.

An  analysis of companion masses revealed a crucial threshold
 round 1.30$\ms\pm0.22\ms$, ~ playing a significant  role in the evolutionary process of DNSs. Below this threshold, the formation via the $ECSNe$ is represented by merging objects characterized by minimal orbital eccentricity and transverse velocity. Note that PSR J1906+0746 is considered a strong candidate for formation through $ECSNe$, even with a companion mass of 1.32\ms.

On the other hand,  the $CC$ model, involving non-merging objects requires a companion mass greater than 1.30\ms. Our analysis suggests that in several systems, the second-born NS likely originated from $ECSNe$, raising the likelihood of a low kick velocity.  This result would leave the  NS provisionally bound post-explosion with low eccentricity. This mechanism can also produce a low-mass NS.

Finally, the conclusions of the present investigation are considered an initial effort to study the physical parameters and distributions of DNSs. It is also a promising step toward understanding the  formation and evolution processes of these systems. We strongly recommend further observations, using more sensitive instruments on future missions (e.g., advanced Virgo detector\footnote{https://www.virgo-gw.eu/}, Einstein Telescope or Cosmic Explorer\footnote{https://www.et-gw.eu/}, LOFAR\footnote{https://www.astron.nl/lofar2-0-newsletter/lofar-development-newsletter-december-2024/}. and FAST\footnote{https://fast.bao.ac.cn/}), which might lead to more discoveries of relativistic DNSs. Furthermore, the planned SKA\footnote{https://www.skao.int/en} is expected to find more DNSs and provide important clues about the nuclear matter compositions of these pulsars.

\acknowledgements We are grateful to Y. Yang for his suggestions that allowed us to improve the clarity of the
original version. Special thanks to B.-S. Liu for helping with some figures.

\bibliography{References}

\end{document}